\documentclass[aps,prd,twocolumn,showpacs,superscriptaddress,nofootinbib,preprintnumbers]{revtex4-2}

\usepackage{style}

\begin{document}

\preprint{FERMILAB-PUB-24-0124-T}
\preprint{MIT-CTP/5703}

 \title{Hitting the Thermal Target for Leptophilic Dark Matter 
 }

\author{Cari Cesarotti \,\orcidlink{0000-0001-5128-7919}}
\email{ccesar@mit.edu}
\affiliation{
Center for Theoretical Physics, Massachusetts Institute of Technology,
Cambridge, MA 02139, USA}


\author{Gordan Krnjaic\,\orcidlink{0000-0001-7420-9577}} 
\email{krnjaicg@fnal.gov}
\affiliation{Theoretical Physics Department, Fermi National Accelerator Laboratory, Batavia, Illinois 60510}
\affiliation{Department of Astronomy and Astrophysics, University of Chicago, Chicago, IL 60637}
\affiliation{Kavli Institute for Cosmological Physics, University of Chicago, Chicago, IL 60637}

\date{\today}

\begin{abstract}
We study future lepton collider prospects for testing predictive models of leptophilic dark matter candidates with a thermal origin.
We calculate experimental milestones for testing the parameter space compatible with freeze-out and the associated collider signals at past, present, and future facilities.
This analysis places new limits on such models by leveraging the utility of lepton colliders. 
At $e^+e^-$ machines, we make projections using precision $Z$-pole observables from $e^+e^-\to \ell^+\ell^- + \slashed{E}$ signatures at LEP and future projections for FCC-ee in these channels.
Additionally, a muon collider could also probe new thermal relic parameter space in this scenario via $\mu^+\mu^- \to X + \slashed{E}$ where $X$ is any easy identifiable SM object. 
Collectively, these processes can probe much all of the parameter space for which DM direct annihilation to $\ell^+\ell^-$ yields the observed relic density in Higgs-like models with mass-proportional couplings to charged leptons.

\end{abstract}

\bigskip
\bigskip
\maketitle

\section{Introduction}
\label{sec:intro}

Thermal freeze-out is the only dark matter (DM) production mechanism that is insensitive to unknown high-energy physics above the scale of the DM mass. 
Over the past several decades, this mechanism has primarily been associated with 
the Weakly Interacting Massive Particle (WIMP) paradigm, which predicts the DM mass to be roughly within 10 GeV -- 10 TeV window \cite{Jungman:1995df}.
While there is a rich experimental program to search for WIMPs, no verified signal has yet been identified, which motivates broadening the search effort outside this mass window and to a wider range of interactions. 

It is well known that high energy colliders are powerful probes of thermal dark matter; indeed, given the interactions strengths that such machines can probe, nearly every DM candidate testable at colliders was in chemical equilibrium in the early universe. 
There are currently strong limits on various dark matter models based on existing LEP \cite{Fox:2011fx}, Tevatron \cite{Bai:2010hh}, and LHC \cite{Kahlhoefer:2017dnp} data.
There are also multiple studies of DM sensitivity at proposed future facilities including the ILC \cite{Dreiner:2012xm}, FCC-ee \cite{Bernardi:2022hny}, FCC-hh \cite{Han:2018wus}, CLIC \cite{Blaising:2021vhh}, CCC \cite{Bai:2021rdg}, CEPC \cite{Liu:2019ogn}, HL-LHC \cite{Cepeda:2019klc}, and various muon collider run options \cite{AlAli:2021let,Han:2020uak,Han:2022ubw,Bottaro:2021snn, Bottaro:2022one}.

\begin{figure*}[t]
    \includegraphics[width=3.4 in]{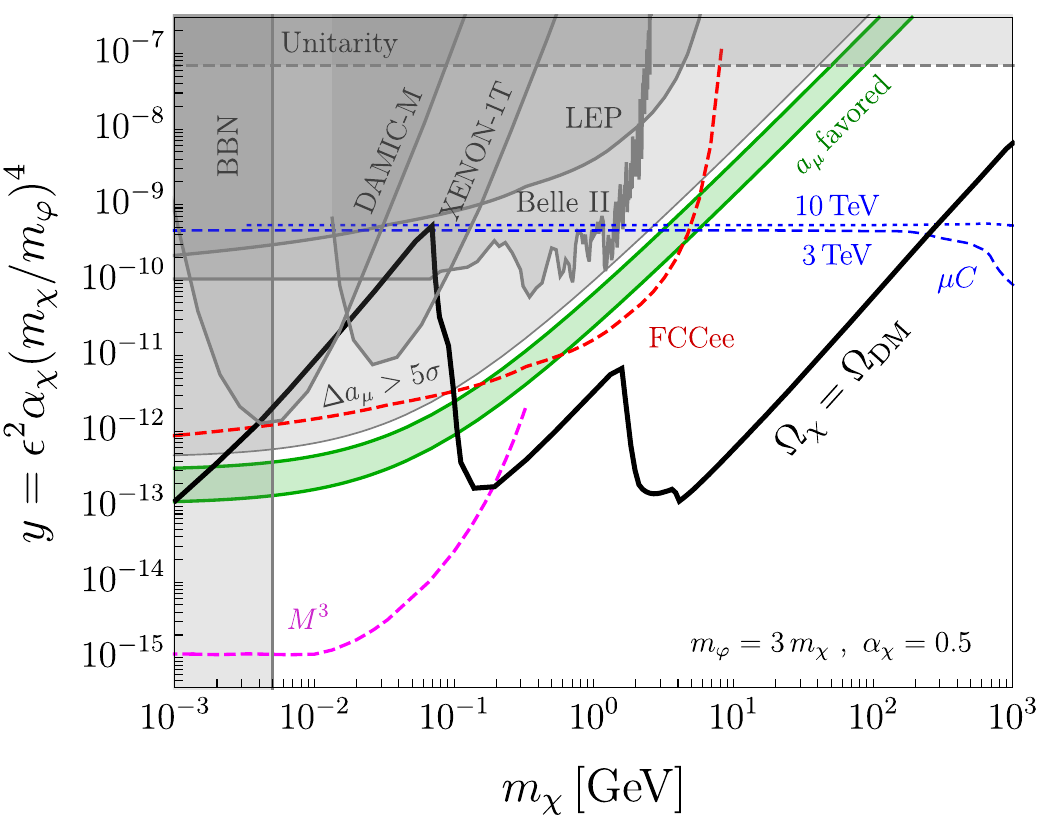}
    \includegraphics[width=3.4 in]{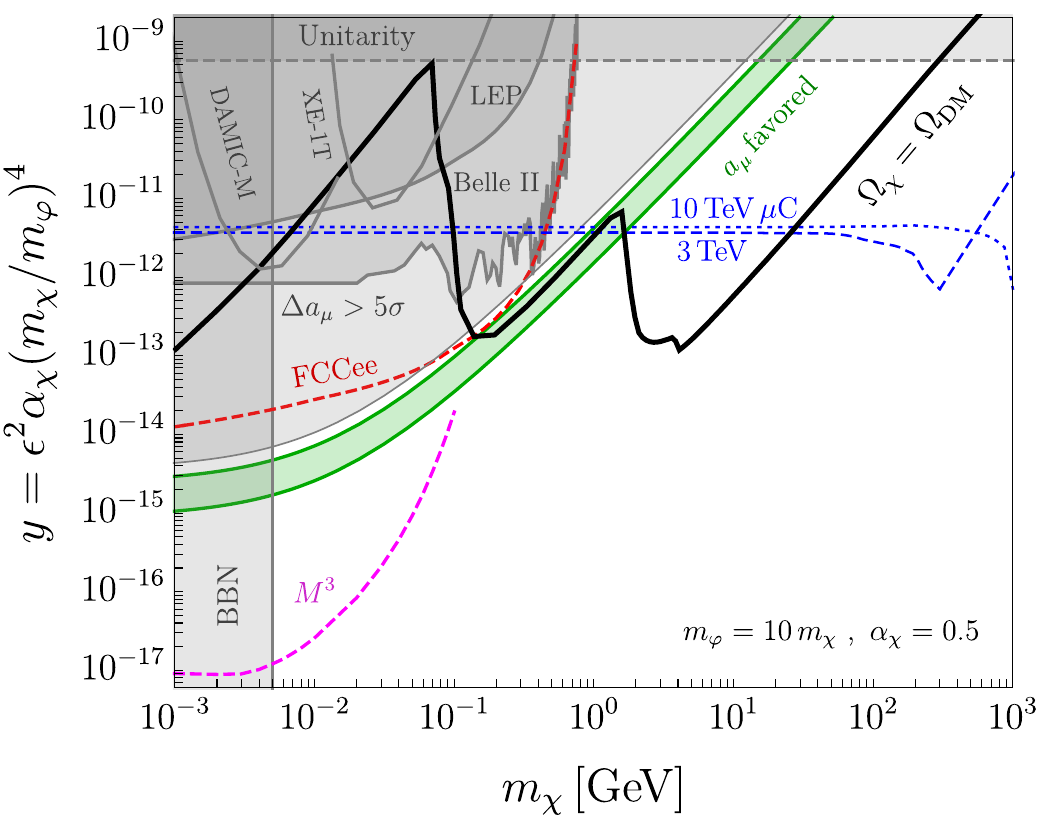} 
    \includegraphics[width=3.4 in]{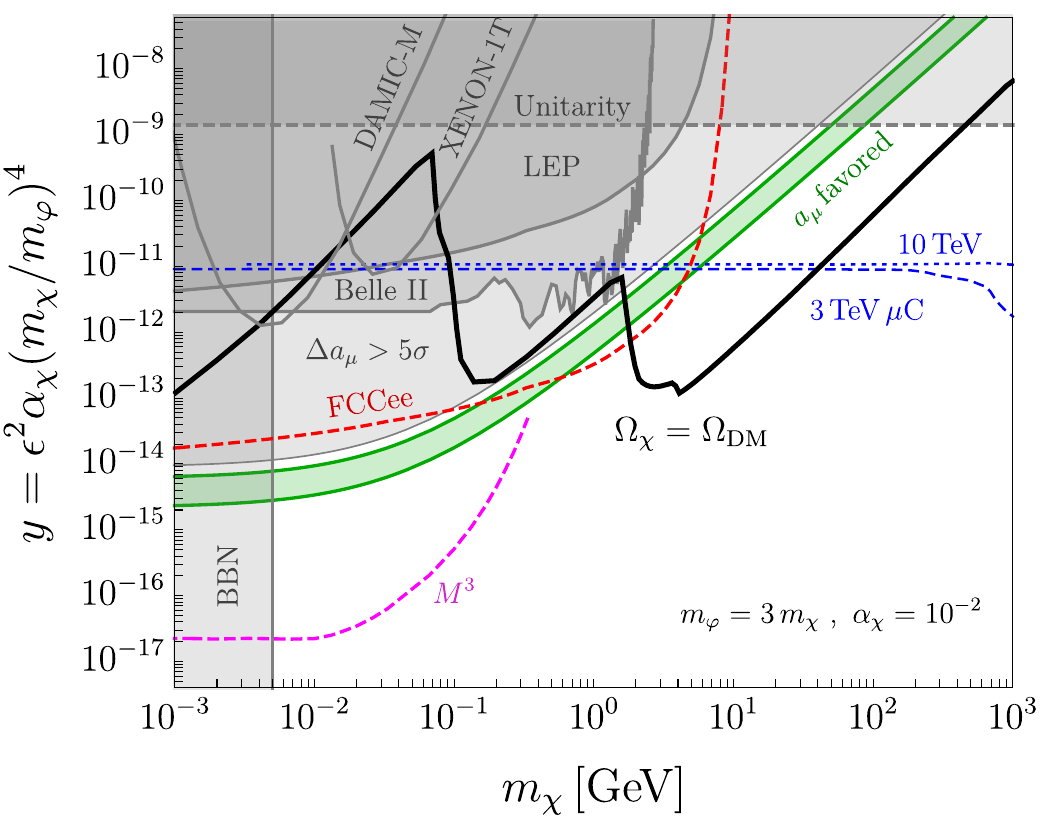}
    \includegraphics[width=3.4 in]{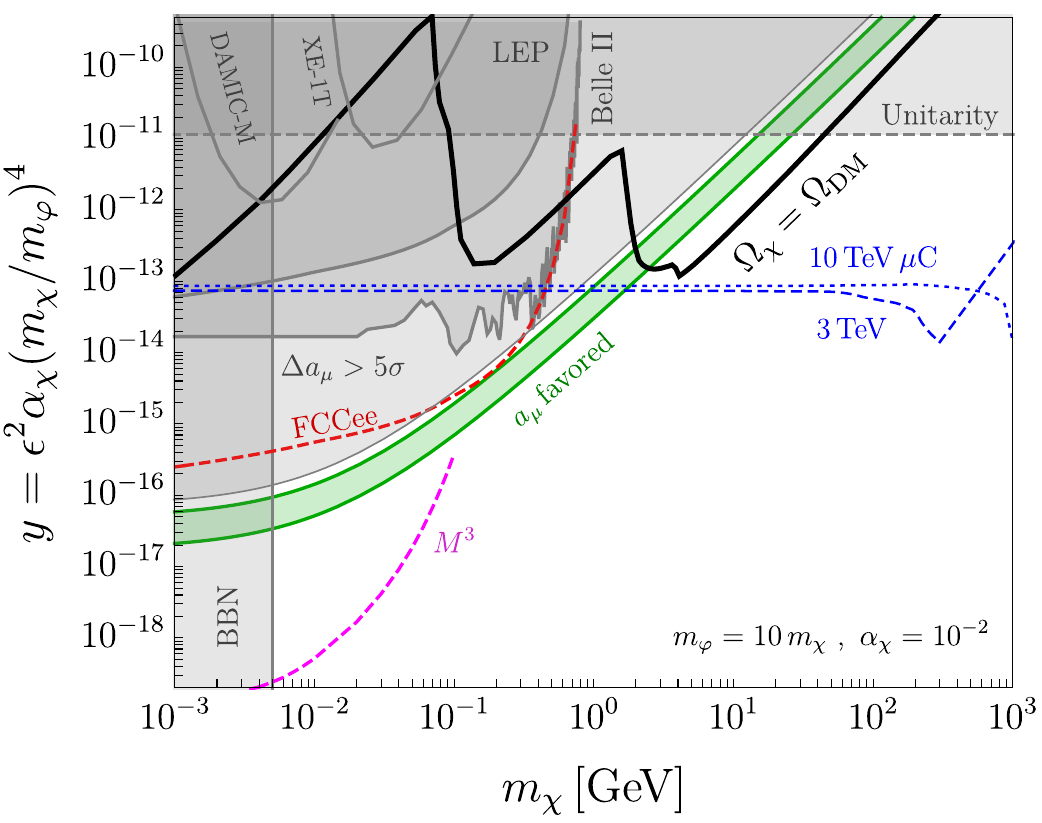}
    \caption{Experimental milestone for $\chi \chi \to \ell^+ \ell^-$ thermal freeze-out (black curve) plotted
    alongside gray-shaded constraints from the muon anomalous magnetic moment $a_\mu$ \cite{Muong-2:2023cdq}, electron recoil direct detection limits from DAMIC-M \cite{DAMIC-M:2023gxo} and XENON-1T \cite{XENON:2019gfn,XENON:2021qze}, existing collider limits from LEP \cite{Drees:2001xw} and Belle II \cite{Belle-II:2022yaw}. The green band indicates the favored parameter space for resolving the $(g-2)_\mu$ anomaly. The four plots show various parameter choices to illustrate a range of phenomenology: the ratio of $m_\varphi/m_\chi = 3, 10$ and the coupling within the dark sector $\alpha_\chi$ = 0.5, $10^{-2}$. The dashed curves are explained throughout Section~\ref{sec:collider}.}
    \label{fig:main}
\end{figure*}

In many of these studies, the most promising DM discovery channel is missing energy in association with a single SM object $X$ (typically a single jet or gauge boson). 
However, if DM couples preferentially to heavy flavor, there may be additional signals that improve upon the mono-$X$ strategy, offering more handles for both signal detection and background rejection.
For example, DM that couples mainly to $b$-quarks, motivated by the galactic center excess, can be probed using $pp \to b\bar b \slashed{E}$ production \cite{Lin:2013sca,Izaguirre:2014vva}.
This strategy can readily be adapted to leptophilic DM that couples primarily to muons and taus.
Production of DM is therefore accompanied by additional charged leptons, whose presence may enhance signal sensitivity.

In this paper we consider the future collider discovery potential for models of leptophilic dark matter with  mass-proportional couplings \cite{Cirigliano:2005ck}, to Standard Model (SM) leptons. 
In the predictive parameter space for this class of models, the scalar particle $\varphi$ that mediates dark-visible interactions is heavier than the DM  $\chi$ ($m_\varphi > m_\chi)$, so that, during freezeout, the process
\be
\label{eq:direct}
\chi \chi \to \varphi^* \to \ell^+ \ell^-
\ee
directly annihilates DM into SM leptons to yield the observed relic density $\Omega_{\chi}h^2 = 0.120 \pm 0.001$ \cite{Planck:2018vyg}. 
Since the cross section for this process depends on the coupling between $\varphi$ and charged leptons, there is a minimum value required for viable freeze-out.
Thus, a thermal origin through the direct annihilation in \Eq{eq:direct} predicts a {\it minimum} signal yield for DM collider production via
\be
\label{eq:new-signals}
e^+e^- \, , \,  \mu^+\mu^- \to \ell^+ \ell^- \varphi ~~,~~
\ee
and there is a corresponding experimental milestone for decisively testing models with this annihilation topology. For $m_\varphi > 2m_\chi$, the mediator decays predominantly to dark matter, so the processes in 
\Eq{eq:new-signals} complement traditional mono-$X$ searches
\be
e^+e^- \, , \,  \mu^+\mu^- \to \, X \varphi \, 
\ee
which are also present in this part of parameter space and included in our analysis of muon collider reach.
 Our study finds that lepton colliders with $e^+e^-$ or $\mu^+\mu^-$ beams are powerful probes this class of models and can test nearly all the remaining parameter space that is compatible with a thermal relic origin. 

 The outline of this work is as follows.
 In Section \ref{sec:theory} we give a theoretical overview for this scenario and calculate the relic abundance, in Section \ref{sec:collider} we present limits and future projections for 
 various lepton collider searches, in Section \ref{sec:other}
 we review existing bounds from previous experiments, and finally in Section \ref{sec:disc} we offer some concluding remarks.

\section{Theory}
\label{sec:theory}

\subsection{Model Description}
Our representative benchmark scenario consists of a  Majorana fermion DM candidate $\chi$ 
with mass $m_\chi$ coupled to a CP-even scalar mediator $\varphi$ with mass $m_\varphi$ via 
\be\label{eq:lag} 
{\cal L}_{\rm int} = -\frac{g_{\chi}}{2}  \varphi \chi \chi 
-  \varphi \!\sum_{\ell = e, \mu,\tau} \! g_\ell \, \bar \ell \ell ~~,~~ g_\ell = g_e \brac{m_\ell}{m_e},
\ee
where $g_{\chi}$ and $g_\ell$ are respectively the Yukawa couplings to DM and charged leptons $\ell$.
Note that the mass proportionality in $g_\ell$ can naturally arise in a Type-III two-Higgs doublet
model supplemented with our singlet scalar $\varphi$ if the latter mass-mixes primarily with the heavy CP-even eigenstate $H$ and has suppressed mixing with the SM-like 125 GeV eigenstate $h$ (see Appendix \ref{sec:2HDM} for a discussion and Ref. \cite{Craig:2013hca} for a review of two Higgs doublet models.).

In order for thermal freeze-out to be predictive and testable with laboratory measurements, the cross section for DM annihilation must depend on the $\varphi$ coupling to the SM, i.e. annihilation cannot occur strictly in the dark sector. 
This requirement is readily achieved by demanding $m_\varphi > m_\chi$, so that the $\chi \chi \to \ell^+ \ell^-$ topology is kinematically allowed for all $\ell$ satisfying $m_\chi > m_\ell$. 
By contrast, if $m_\chi > m_\varphi$, the dominant channel for annihilation will be  secluded annihilation via $\chi\chi \to \varphi\varphi$, which only depends on $g_\chi$ and there is 
no minimum requirement on $g_\ell$ so long as $\chi$ thermalizes with the SM in the early universe.
Note that this secluded channel yields an $s$-wave annihilation cross section, which is excluded for thermal relics below $\sim$ 20 GeV (see Sec. \ref{sec:other} below). 
Thus, for the reminder of this work, we only consider the case where $\varphi$ is heavier than $\chi$.  

Furthermore, since experimental limits on $g_\ell$ are generically stronger than those on $g_\chi$, we will work in the regime where, in the $m_\varphi > 2m_\chi$ regime, the scalar will decay predominantly via $\varphi \to \chi \chi$. 
Thus, we are interested in signatures involving heavy lepton production in association with missing energy rather than $\varphi$ decaying back to SM particles as multi-lepton final states are well constrained at colliders.

We also note that similarly predictive models with Higgs-mixed mediators, whose couplings to all SM fermions is proportional to their massses \cite{Krnjaic:2015mbs}. However, in the parameter space where DM annihilates directly to SM particles, this model faces stringent limits from various laboratory measurements (most notably rare meson decays) which robustly exclude thermal freeze-out for all DM masses. However, since most of these limits are based on the large top Yukawa coupling, our leptophilic scenario remains viable across a wide range of model parameters.

\subsection{Cosmological Production}
In the early universe for $T \gg m_\chi$, we assume the DM is in thermal equilibrium with the SM. 
As the universe cools below $T \sim m_\chi$, the DM population becomes Boltzmann suppressed and eventually the abundance of DM freezes out.
We calculate the $\chi$ relic density by solving the Boltzmann equation 
\be\label{eq:boltz}
\dot n_\chi + 3 H n_\chi = -\langle \sigma v \rangle [ n_\chi^2 - (n_\chi^{\rm eq})^2],
\ee
where $H = 1.66 \sqrt{g_\star}T^2/M_{\rm Pl}$ is approximately the Hubble expansion rate, $g_\star$ is the effective number of relativistic species, $M_{\rm Pl}$ is the Planck mass, and $n_\chi^{\rm eq}$ is the equilibrium $\chi$ number density.
For $m_\varphi > m_\chi$, the thermally averaged annihilation cross section satisfies 
\be
\label{eq:svtot}
\langle \sigma v \rangle = \sum_{\ell = e,\mu,\tau}  \langle \sigma  v\rangle_{\chi \chi \to \ell \ell} ,
\ee
where the sum is over all kinematically open channels ($m_\chi > m_\ell$). 

To obtain the thermal average in \Eq{eq:svtot}, we numerically calculate the total annihilation cross section as a function of Mandelstam variables and follow the prescription from Ref. \cite{Gondolo:1990dk}:
\be
\label{eq:svTA}
\langle \sigma v \rangle_{\chi \chi \to \ell \ell}
= \frac{1}{N}
\int_{4m_\chi^2}^\infty \! ds\,
\sigma(s) (s-4 m_\chi^2)\sqrt{s} K_1  \!
\left(  \!\frac{\sqrt{s}}{T} \right)\!,\,
\ee
where 
$N = 8  m^4_\chi T   K^2_2\left( \frac{m_\chi}{T} \right)$ is a normalization factor, and 
$K_{1,2}$ are modified Bessel functions of
the first and second kinds. For
each $\chi\chi \to \ell^+ \ell^-$ channel, the total annihilation cross section is 
\be
\sigma(s) = \frac{ g_\ell^2 g_\chi^2 (s-4m_\chi^2)^{3/2} }{ 16 \pi [  (s-m_\varphi^2)^2 + m_\varphi^2\Gamma_\varphi^2  ]}
\sqrt{1-\frac{4m_\chi^2}{s}}~,
\ee
where $\Gamma_\varphi$ is the total $\varphi$ decay width,
and we neglect subleading corrections of order $m^2_\ell/s$.

Although our numerical results utilize the full thermally averaged cross section in \Eq{eq:svTA}, it is useful to consider the form of $\sigma v$ in the non-relativistic limit prior to performing the thermal average. In this regime, each channel contributes 
\be
\label{eq:sigmav}
 \sigma  v_{\chi \chi \to \ell \ell}  \approx \frac{g_\chi^2 g_\ell^2 m_\chi^2  v^2}{8\pi (m_\varphi^2 - 4m_\chi^2)^2 } \propto g_\chi^2 g_\ell^2 \brac{m_\chi}{m_\varphi}^4 \frac{1}{m_\chi^2},
\ee
where $ v$ is the relative velocity between initial state $\chi$ particles. 
Matching the convention in Refs. \cite{Izaguirre:2015yja,Berlin:2018bsc}, we define the dimensionless variable 
\be
\label{eq:y}
y \equiv  \epsilon^2 \alpha_\chi \brac{m_\chi}{m_\varphi}^4, ~~  \epsilon \equiv g_e/e~,
\ee
where $\alpha_\chi \equiv g_\chi^2/(4\pi)$.
We normalize the parameter $\epsilon$ in units of the electron charge $e$ in analogy with the familiar kinetic-mixing parameter in models with dark photon mediators \cite{Fabbrichesi:2020wbt}.
This quantity is useful because, away from the fine tuned resonance region $m_\varphi \approx 2m_\chi$, there is a one-to-one relationship between $\sigma v$ and $y$ independently of the $m_\varphi/m_\chi$ or  $g_\chi/g_\ell$ ratios; here the relic density for each value of $m_\chi$ is simply set by the product that defines $y$.
In Fig. \ref{fig:main}, the black curves in the $y$ vs. $m_\chi$ plane represent the parameter points for which $\chi\chi \to \ell\ell$ annihilation yields the observed dark matter abundance through thermal freeze-out.

Importantly, the non-relativistic expansion in \Eq{eq:sigmav} also shows that this process is $p$-wave, so the annihilation rate falls precipitously as the universe expands and the temperature cools.
As discussed below in Section \ref{sec:other}, this feature sharply suppresses the annihilation to SM particles during and after the epoch of recombination.

\section{Collider Searches}
\label{sec:collider}
In this section we present the lepton collider signatures that the model in Section \ref{sec:theory} predicts.
Here we put new limits on this scenario using existing data from LEP and Belle II, and we make future predictions for FCC-ee and a future muon collier. 
The results are summarized in Fig.~\ref{fig:main}, with the excluded regions shaded gray and predictions for future experiments given in dashed lines.

\subsection{Belle II}
Since $\varphi$ has sizeable couplings to heavy flavor, it can be produced as final state radiation at experiments where 2nd and 3rd generation dilepton pairs are copiously produced, such as at $B-$factories via 
\be
e^+e^- \to \mu^+\mu^- \varphi,~~ (\varphi\to \chi  \chi),
\ee
where the dark matter $\chi$ in the final state produces missing energy.
This signal has an irreducible background from 
$e^+e^- \to \mu^+ \mu^- \bar \nu \nu$
from the missing energy in the final state neutrinos, as well as reducible backgrounds from double
bremsstrahlung
$e^+e^- \to \mu^+ \mu^-  \gamma \gamma$. 
The Belle II collaboration has placed constraints on an $L_\mu - L_\tau$ gauge boson through this search channel \cite{Belle-II:2022yaw}.
We recast these limits for our scalar scenario by using \texttt{MadGraph5\_aMC@NLO} \cite{Alwall:2014hca} to calculate the cross section for $e^+e^- \to \mu^+\mu^-\varphi$ production and rescale the limits for the $L_\mu-L_\tau$ vector model with the corresponding cross section for vector particles. 
We neglect any signal acceptance differences between scalars and vector and vector production. 

We note that, in principle, $B$-factories can be sensitive to $\varphi$ production 
as final state radiation in $e^+ e^- \to \tau^+ \tau^- \varphi$ channels (considered
below for higher energy lepton colliers).
However, despite the large $\tau$ enhancement in this channel, to date, there has been no dedicated search for such a process at these facilities for $\varphi$ decaying invisibly\footnote{ However, there are strong BABAR limits on $e^+e^- \to \tau^+\tau^-$ followed by a $\varphi \to \mu^+\mu^-$ decay \cite{BaBar:2020jma}}. 
We leave a detailed study of such signals for future work. 

\begin{figure}[t]
\hspace{-0.5cm}
    \includegraphics[width=1.7 in]{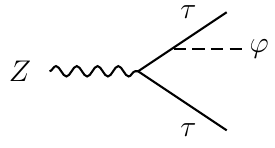}
    \medskip
    \caption{Representative Feynman diagram of a rare $Z \to \tau^+ \tau^- \varphi$ decay that generates
    our signal of interest at LEP and FCC-ee. Here the $\varphi \to \chi \chi$ produces
    additional missing energy relative to SM $Z \to \tau^+ \tau^-$ decays.}
    \label{fig:feyn}
\end{figure}

\subsection{LEP}
It has long been established that the LEP and LEP-2 $e^+e^-$ colliders at CERN \cite{Drees:2001xw} have impressive sensitivity to dark matter production through mono-$\gamma$ and missing energy $\slashed{E}$ signatures \cite{Fox:2011fx}. 
While these canonical signatures can also arise in our scenario from the couplings in \Eq{eq:lag}, they are not the dominant channel as the production cross section for $e^+e^- \to \gamma \varphi \to \gamma \slashed{E}$ is suppressed by the small electron coupling $g^2_e$. 
Since the $\varphi$ coupling to the $\tau$ is
enhanced by a factor of $(m_\tau/m_e)^2 \sim 10^6$, here we instead consider the signature
\be
e^+ e^- \to \tau^+ \tau^- \varphi ~~,~~ (\varphi \to \chi \chi),
\ee
where the invisibly decaying mediator is produced as final state radiation and predicts deviations within well-measured SM uncertainties.
Since SM $\tau$ decays always produce missing energy via $\nu$, our signal process will be constrained by any previous measurement of $Z$ decays that result in $\tau$ production \cite{ALEPH:2005ab}. 
Up to small differences in geometric acceptance and final state $\tau$ kinematic distributions, $Z\to \tau^+ \tau^-$ production is experimentally indistinguishable from $Z\to \tau^+ \tau^- \varphi$ for $\varphi$ decaying invisibly.

The one of the primary motivations for a high-energy electron-positron collider is a run at the $Z$ pole where $\sqrt{s} \approx m_Z$, so the $Z$ production cross section is resonantly enhanced and with copious production, precision measurements can be made. 
Therefore, LEP is mainly sensitive to $\varphi$ production through modifications to the well-measured properties of the $Z$. 
Of particular interest is the signal process depicted in Fig. \ref{fig:feyn}
\be
Z \to \tau^+\tau^- \varphi, ~ (\varphi \to \chi \chi)~,
\ee
where the $\varphi$ decay to dark matter yields additional missing energy in the final state. 
In Appendix \ref{sec:zdecay} we calculate the partial width for this 3-body process. 
From LEP,  we obtain a limit on $\varphi$ production from the measurement of the di-tau partial width \cite{ParticleDataGroup:2020ssz}
\be
\Gamma({Z\to \tau^+\tau^-}) =   84.08 \pm 0.22 \, \rm MeV.
\ee
From the uncertainty we extract a 2$\sigma$ limit on the BR($Z\to \tau^+\tau^- \varphi$), we find the value of $g_\tau$ at each value of $m_\varphi$ that satisfies 
\be
\Gamma(Z \to \tau^+ \tau^- \varphi)  =  2 (0.22 \, \rm MeV)~,
\ee
 and translate this into a constraint on $y$ for a given choice of $\alpha_\chi$ and $m_\chi/m_\varphi$ ratio. 

We also include a highly simplified estimate for the acceptance of $Z \rightarrow \tau \tau \varphi$ events due to analysis cuts on kinematics, which is discussed in Appendix~\ref{sec:taus}. 
 In Fig. \ref{fig:main} this bound 
is shaded in gray and labeled LEP. 

\subsection{Future Circular Collider}
The Future Circular Collider (FCC) at CERN is a proposed high-energy collider to succeed the LHC at CERN.
The first stage of the project is the FCC-ee, an $e^+e^-$ collider to run at similar run configurations of LEP but with much greater luminosity, which aims to take data in the early 2040s \cite{Agapov:2022bhm,Bernardi:2022hny, FCC:2018byv, FCC:2018vvp, FCC:2018evy}. 
As part of its overall physics program, FCC-ee will run at the $Z$-pole and produce 
over one trillion $Z$ bosons for precision electroweak measurements (called the \textit{Tera-Z run}), thereby exceeding the previous LEP record of approximately 17 million \cite{Drees:2001xw}. 
This allows for a diverse physics program that involves precision measurements of $Z$ decays (e.g. Refs.~\cite{Kamenik:2023hvi, Ho:2022ipo})

To make sensitivity projects of the FCC-ee at the $Z$ pole, we assume statistical uncertainties dominate the FCC-ee error budget and rescale the uncertainty on the $Z$ width measurement by $\sqrt{N_\text{LEP}/N_\text{FCC-ee}}$ for $N_X$ the total number of $Z$ bosons produced.
With this estimation we predict sensitivities to $Z$-boson branching ratios comparable to what is stated in Ref.~\cite{Agapov:2022bhm}.
In Fig. \ref{fig:main} we show projections of the improvement of FCC-ee over LEP with the red dashed curve.

\subsection{Future Muon Collider} 
Another possible option for a future high-energy collider is a multi-TeV muon collider (MuC) \cite{Budker:1969cd, earlyMuon:1996, osti:1156195, Mohayai:2018rxn, Delahaye:2014vvd, Boscolo:2018ytm}. 

Due to their moderate mass and fundamental nature, muons can theoretically be accelerated to collide at $\sqrt{\hat{s}} \sim  10 \text{ TeV}$ energies without insurmountable loses due to synchroton radiation. 
Additionally, as muons couple predominantly to electroweak gauge bosons and do not produce enormous hadronic backgrounds, a high-energy muon collider could provide a window into both the 10-TeV frontier as well as precision electroweak measurements within the next 30 years \cite{Barger:1997yk, AlAli:2021let, Han:2020uid, Han:2021lnp, Buttazzo:2018qqp, Buttazzo:2020uzc, Forslund:2023reu, Ruhdorfer:2023uea}. 
While this technology needs a full demonstrator program to prove feasibility, consider theoretical, experimental, and accelerator work has already gone into its design (see Ref.~\cite{Accettura:2023ked} for a recent summary), and its construction is already partially funded. 

As a muon collider is still in its very early stages of planning, the run configurations have not yet been fixed and a multitude of possible center-of-mass energies are considered.
The guiding principle for projections is that the total luminosity must scale with $s$ such that we maintain sensitivity to $s$-channel processes at high energies.
However, due to the fact that muons decay, higher-energy muon beams will have a higher luminosity from time dilation effects. 

In this paper, we utilize the high energy of the muon collider to probe complementary parameter space to what can be done at the FCC-ee. 
We consider the process $\mu^+ \mu^- \to \gamma \varphi \to \gamma \slashed{E}$ with a mono-$\gamma$ signature. 
Because of the kinematics of the process, the $\gamma$ will have energy roughly $E_\gamma \sim \sqrt{s}/2$ depending on the width of the $\varphi$ and energy resolution $\sigma_E$. 
The primary SM background for this process is $\mu^+ \mu^- \to \gamma \bar{\nu}\nu$.\footnote{We also considered the 3$\gamma$ background, but as two photons, one with $E\sim$ TeV have to be missed, this background is comparatively negligible.}
We simulate both the signal and background events in \texttt{MadGraph5\_aMC@NLO} \cite{Alwall:2014hca}. 
To mitigate this background, we utilize the fully-constrained kinematics of the 2-body final state of the signal and cut on the energy of the visible photon:
\begin{equation}
E_\gamma \in \left( E_0 (1 - 2 \sigma_E), E_0 (1+2\sigma_E) \right)
\end{equation}
where $E_0 \equiv \frac{E_{\rm CM}^2 - m_\varphi^2}{2 E_{\rm CM}}$ and $\sigma_E$ is the energy resolution we fix nominally at $\sigma_E = 3$\%. 
We also take a conservative photon reconstruction efficiency of 98\%, although we expect this number for TeV photons might be higher.
The search strategy in this channel is generic as long as the mediator decays invisibly via $\varphi \to \chi \chi (\slashed{E})$ and the other SM particles $X$ recoil in a diametrically opposed direction to the missing energy and momentum in the final state. 
A similar strategy with partially visible decays was considered in Ref.~\cite{Capdevilla:2021kcf}.

The dashed blue curve in Fig. \ref{fig:main} is a sensitivity projection for $\mu^+ \mu^- \to X \varphi \to X \slashed{E}$ at  a 3 TeV muon collider concept with 1 ab$^{-1}$ of integrated luminosity.\footnote{We do not consider a lower energy center of mass, like the Higgs pole, as higher-energy configurations have a broader physics program and can even produce upwards of $10^6$ to $10^8$ Higgs.}

\section{Other Searches}
\label{sec:other}

In this section we describe the other constraints on this model from precision measurements, cosmology, and direct detection experiments. 

\subsection{Muon Anomalous Magnetic Moment}
The  Brookhaven E821 \cite{Muong-2:2006rrc} and  Fermilab Muon $g-2$ \cite{Muong-2:2023cdq} experiments
have measured the anomalous magnetic moment of the muon, resulting in a world average of
\be
a_\mu^{\rm exp} =
  116 592 059 (22) \times   10^{-11}~~.
\ee
The SM prediction based on R-ratio calculations is \cite{Aoyama:2020ynm}
\be
a_\mu^{\rm SM} = 116591810(43) \times   10^{-11}~~,
\ee 
which differs from the observed value by
\be
\Delta a_\mu = (249 \pm 48) \times 10^{-11},
\ee
and thereby exceeds the 
nominal 5$\sigma$ benchmark for a new physics discovery. 

However, a lattice
calculation from the BMW collaboration extracts a 
SM prediction for $a_\mu$ that is closer to the observed 
value and in tension with the R-ratio result. Additional lattice results will be necessary to independently verify the BMW result and 
future R-ratio data will clarify the tension between BABAR, KLOE, and CMD-3 measurements near $\sqrt{s} \approx$ 1 GeV \cite{CMD-3:2023alj,Aoyama:2020ynm,Muong-2:2023cdq}. Thus, in our analysis, we remain agnostic about whether the discrepancy is due to new physics. 

Since the leptophilic mediator $\varphi$ couples to muons, at one loop it modifies $a_\mu$ by 
\be
\Delta a_\mu = \frac{g_\mu^2}{16\pi^2} \int_0^1 dz \frac{
 (1-z)(1-z^2)
}{ (1-z)^2 + r^2 z} ,
\ee
where $r = m_\varphi/m_\mu$, which constrains the thermal relic parameter space shown in Fig. \ref{fig:main}. The gray region labelled ``$\Delta a_\mu > 5 \sigma$" is where model parameters make the tension with observation worse by 5 standard deviations. For parameters within the green band labeled ``$a_\mu$ favored," the $\varphi$ coupling to muons accounts for the discrepancy between the measured world average and the $R$-ratio prediction within the Standard Model assuming the difference is due to new physics.

\subsection{Fixed Target Experiments}

For the sub-GeV DM mass range, proposed fixed target 
accelerators are powerful laboratory probes of light weakly coupled particles \cite{Krnjaic:2022ozp}. The mediator $\varphi$ can be radiatively produced
in muon-nucleus fixed target interactions via 
\be
\mu N \to \mu N \varphi, ~~(\varphi \to \chi\chi)~,
\ee
where $N$ is a target nucleus.  For $m_\varphi > m_\mu$, the radiated $\varphi$ typically acquires a large fraction of the incident beam energy, which serves as a trigger for a dark matter search if other visible particles produced in $\mu$-$N$ scattering can be vetoed. In Fig. \ref{fig:main} we show projections for the proposed $M^3$ experiment \cite{Kahn:2018cqs,Capdevilla:2021kcf} based on this search strategy. Comparable projections could also be computed for the CERN NA64$\mu$ experiment which performs a similar search at higher energy \cite{Andreev:2024sgn}, but translating the publicly available projections into the scalar model considered here is beyond the scope of this work. 
 
\subsection{Direct Detection}

\subsubsection{Electron Recoils}

Since the mediator $\varphi$ couples to electrons, this scenario
is constrained by direct detection experiments sensitive to scattering off electrons. The 
non-relativistic cross section for $\chi e \to \chi e$ scattering is
\be
\sigma_{\chi e} = \frac{ g_\chi^2 g_e^2 \mu^2_{\chi e} }{\pi[(\alpha m_e)^2+m_\varphi^2]^2} \to \frac{ g_e^2 g_\chi^2 \mu_{e\chi}^2}{\pi m_\varphi^4} \approx \brac{4 e^2 m_e^2}{m_\chi^4} y,~~
\ee
where $\mu_{\chi e}$ is the $\chi$-$e$ reduced mass and
in the last step we took the the $m_e \ll m_\varphi, m_\chi$ limit, where
$\sigma_{\chi e} \propto y$ and 
the direct detection cross section has the same parametric dependence as the $\chi \chi \to e^+ e^-$ annihilation cross section from \Eq{eq:sigmav}. Although the electron scattering sensitivities various experiments are comparable in this mass range, in Fig. \ref{fig:main} we show the current best limits from the DAMIC-M \cite{DAMIC-M:2023gxo} and XENON-1T \cite{XENON:2019gfn,XENON:2021qze} experiments.

\subsubsection{Nuclear Recoils}
Although $\varphi$ has no tree-level coupling to quarks,  at one-loop there is induced mass-mixing between $\varphi$ and the SM Higgs, which yields the interactions
\be
\label{eq:lag-mix}
{\cal L}_{\rm int} \supset \theta y_q \varphi \bar q q \equiv g_q \varphi \bar q q,
\ee
where $q$ is any SM quark, $y_q = m_q/v_h$ is its SM Yukawa coupling to the Higgs field, and $\theta$ is the $\varphi$-$h$ mixing angle. The leading
contribution to this mixing is given by integrating out the $\tau$ loop $\theta \sim g_\tau y_\tau/(16 \pi^2)$.
 Based on this mixing, the elastic spin-independent direct-detection cross section for 
 $\chi$-$n$ scattering is 
\be
\label{eq:sigma-nuc}
\sigma_{\chi n} =  \frac{g_\chi^2 g^2_{n}  \mu^2_{\chi n}}{\pi m_{\varphi}^4} ,
\ee
where $n$ is a target nucleon, $\mu_{\chi n}$ is the DM-nucleon reduced mass, and the induced $\varphi$ nucleon coupling $g_{n} \propto \theta$ can be written in terms of the mediator couplings to quarks in \Eq{eq:lag-mix} following the procedure in Ref.  \cite{Shifman:1978zn} -- see also the appendix of Ref. \cite{Cirelli:2013ufw} for a modern treatment. 
Writing \Eq{eq:sigma-nuc} in terms of 
$y$ from \Eq{eq:y} 
\be
y \approx  10^{-8}\brac{\sigma_{\chi n}}{10^{-46} \, \rm cm^2} \brac{m_\chi}{20 \, \rm GeV}~.
\ee
Thus, for cross sections of order  $\sigma_{\chi n} \approx 10^{-46}$ cm$^2$ at the maximum sensitivity of 
XENONnT \cite{XENON:2023cxc} near $m_\chi \approx 20$ GeV, direct detection
does not meaningfully constrain the 
thermal relic values of $y$ as shown in the black curve of Fig. \ref{fig:main}. 
A similar argument can be used to show that the two-loop
QED scattering processes lead to similar results. 
Thus, we omit nuclear recoil limits 
through loop induced elastic scattering 
in Fig. \ref{fig:main}. However, we note that if the Higgs-mixing parameter $\theta$ is enhanced beyond the $\tau$ loop that we use as a reference value (e.g. from additional tree level contributions), the model dependent constraint from this process could be competitive.

\subsection{Cosmology}

\subsubsection{BBN Dark Radiation}
For $m_\chi \lesssim$ MeV, the $\chi$ freeze-out 
occurs around the epoch of Big Bang Nucleosynthesis (BBN) and can modify light chemical yields relative to SM predictions. In this light mass range, the 
DM has a large thermal abundance at the onset of BBN, which modifies the Hubble expansion rate and affects the timing of BBN events
 \cite{Boehm:2013jpa,Nollett:2013pwa}. Using the latest nuclear cross sections and data on the deuterium fraction, BBN observables require $m_\chi \gtrsim$ 5 MeV for Majorana DM candidates that transfer entropy to photons upon freeze-out \cite{Krnjaic:2019dzc} and this defines the leftmost regions in the panels of Fig. \ref{fig:main}

\subsubsection{CMB Energy Injection}
\label{sec:cmb}
Thermal DM below the GeV scale faces stringent bounds
from CMB energy injection from out-of-equilibrium 
DM annihilation just after recombination
\cite{Finkbeiner_2012}. For an $s$-wave annihilation 
cross section, thermal WIMPs are excluded below $\sim$ 20 GeV
if their annihilation yields visible energy into the CMB \cite{Planck:2018vyg}.
However, in our scenario the annihilation cross section in \Eq{eq:sigmav} is
$p$-wave and therefore suppressed by $v^2$ in the non-relativistic
regime. Thus, the CMB limit from Planck does not constrain the thermal relic parameter space for this model.

\section{Discussion}
\label{sec:disc}
In this paper we have 
studied a predictive leptophilic DM scenario in which the thermal relic abundance arises from direct annihilation to SM particles via $\chi\chi\to \ell^+\ell^-$ with mass-weighted couplings to charged leptons.
Since this scenario has no tree-level couplings to quarks, LHC and nuclear-recoil direct detection signals are suppressed compared to traditional WIMP DM models.  
We have set new limits on this model based on various probes, including measurements of the muon anomalous magnetic moment, electron recoil direct detection, $B$-factory production, and $Z$-pole measurements at LEP. 
Collectively, these limits leave much of the thermal relic parameter space unexplored, particularly for DM masses above the GeV scale where the relic density arises predominantly through $\chi\chi \to \tau^+\tau^-$ annihilation in the early universe. 

Future accelerators will play a key role in probing the remaining  thermal relic parameter space in this scenario. In the sub-GeV mass range, the proposed fixed-target missing-momentum experiment $M^3$ can comprehensively cover the full thermal target and complement ongoing efforts currently underway with the NA64$\mu$ experiment. At higher mass, the proposed FCC-ee  $e^+e^-$ collider can probe large regions of viable parameter space with a Tera-$Z$ run at the $Z$-pole, which can improve sensitivity to $Z \to \tau^+\tau^-$ decays, currently limited by the luminosity of the legacy LEP data set.  And at the largest masses compatible with a viable thermal history $m_\chi \sim $ TeV in this framework, a future TeV-scale muon collider can probe nearly all of the remaining freeze-out parameter space with mono-$\gamma/Z$ searches.

\section*{Acknowledgements}
We would like to thank Andr\'e de Gouv\^ea, Rodolfo Capdevilla, Bertrand Echenard, Maxim Pospelov, Jesse Thaler, Dario Buttazzo, Simone Pagan Griso, and Andrea Wulzer for helpful conversations.
C.C. is supported by the U.S. Department of Energy (DOE) Office of High Energy Physics under Grant Contract No. DE-SC0012567. Fermilab is operated by the Fermi Research Alliance, LLC under Contract DE-AC02-07CH11359 with the U.S. Department of Energy. This material is based partly on support from the Kavli Institute for Cosmological Physics at the University of Chicago through an endowment from the Kavli Foundation and its founder Fred Kavli.

\newpage
\appendix
\onecolumngrid

\section{Realization in a Two Higgs Doublet Model}
\label{sec:2HDM}

The mass weighted couplings in our scenario can arise in a type-III two Higgs (2HDM) doublet model with an additional CP even scalar singlet $S$ that mixes with the SM-like $h$ and heavy Higgs state $H$. In this realization of a 2HDM, before electroweak symmetry breaking, the Yukawa interactions can be written
\be
{\cal L} \supset \lambda_u H_1 Q \bar  u + \lambda_d Q H^\dagger_1 Q \bar d + \lambda_\ell H_2 L \bar e ~,
\ee
where $H_{1,2}$ are electroweak doublets with quantum numbers $2_{\pm 1/2}$ under $SU(2)_L \times U(1)_Y$, so $H_1$ gives mass to the quarks and $H_2$ gives mass to charged leptons. Here we use two-component Weyl fermion notation for the quark and lepton fields, where $Q, L$ are electroweak doublets and $\bar u, \bar d, \bar e $ are $SU(2)_L$ singlet Dirac partners for the up-type, down-type, and charged lepton fields, respectively.

Following  the conventions in
Ref. \cite{Craig:2013hca},   after electroweak symmetry breaking, the vacuum expectation values are $\langle H_i \rangle = v_i$. We define the ratio $\tan \beta \equiv v_2/v_1$ and 
diagonalize the CP even mass eigenstates with the rotation matrix
\be
\begin{pmatrix}
\sqrt{2}\, {\rm Re}(H_1) - v_1  \\
\sqrt{2} \, {\rm Re}(H_2) - v_2  \\
\end{pmatrix}
=
\begin{pmatrix}
~~\cos \alpha & \sin\alpha \\
-\sin \alpha & \cos\alpha \\
\end{pmatrix}
\begin{pmatrix}
h  \\
H  
\end{pmatrix},
\ee
where $h$ is the SM-like Higgs mass eigenstate 
and $H$ is the heavy CP-even state. In 
the mass eigenbasis, the $h,H$ couplings 
to fermions and electroweak gauge bosons (denoted by $V$) are given in Table \ref{tab:couplings}

\begin{table}[t]
\begin{center}
\begin{tabular}{|c|c|c|c|c|} \hline &  $\lambda_{\rm 2HDM}/\lambda_{\rm SM}$ \\ \hline
$hVV$ &   $\sin(\beta - \alpha)$  \\
$h Q \bar u $ & $\sin(\beta - \alpha) +  \cos(\beta - \alpha)/ \tan \beta$  \\
$h Q \bar d$ & $\sin(\beta - \alpha) +  \cos(\beta - \alpha)/ \tan \beta$ \\
$h L\bar  e$ & \hspace{-0.25cm} $\sin(\beta - \alpha) -  \tan \beta \cos(\beta - \alpha) $  \\ \hline
$HVV$ &  $\cos(\beta - \alpha)$   \\
$H Q \bar u$ &  $\cos(\beta - \alpha) - \sin(\beta - \alpha)/ \tan \beta$  \\
$H Q \bar d$ &  $\cos(\beta - \alpha) -  \sin(\beta - \alpha)/ \tan \beta$ \\
$H L\bar  e$  & \hspace{-0.25cm} $\cos(\beta - \alpha) + \tan \beta \sin(\beta - \alpha)$\\\hline
\end{tabular}
\caption{Tree level couplings for CP even eigenstates $h$ and $H$ in a Type III 2HDM normalized to their values in the SM. For large $\tan \beta \gg 1$ in the alignment limit where $\sin(\beta- \alpha) \to 1$, $h$ recovers the SM like couplings and $H$ becomes lepophilic with mass proportional Yukawa couplings. 
\label{tab:couplings}}
\end{center}
\end{table}%

To realize our scenario in \Eq{eq:lag} at low energy, we add a singlet scalar $S$ and introduce super-renormalizable  Higgs-portal couplings 
\be
V(H_1, H_2, S) = S \left( \mu_{11} H_1^\dagger H_1 + \mu_{12} H_1^\dagger H_2 +\mu^*_{12} H_2^\dagger H_1 + \mu_{22} H^\dagger_2 H_2 \right)  ~ ,
\ee
where $\mu_{ij} \ll v_h$ are dimensionful mixing parameters. Writing this expression in terms of mass eigenstates in the electroweak breaking vacuum, we have 
\be
{\rm Re}(H_1)  = \frac{ h \cos\alpha   + H \sin \alpha  + v_1 }{\sqrt{2}}~~,~~
{\rm Re}(H_2)  = \frac{  - h \sin\alpha   + H \cos \alpha  + v_2 }{\sqrt{2}}~,
\ee
so the mixed potential involving $S$ becomes 
\be
V(h, H, S) &=& \frac{S}{2} \biggl[ \mu_{11}  \left( h \cos\alpha   + H \sin \alpha  + v_1 \right)^2
 + 2 {\rm Re (\mu_{12}) } 
  \left( h \cos\alpha   + H \sin \alpha  + v_1 \right)
\left(- h \sin\alpha   + H \cos \alpha  + v_2\right) 
\nonumber \\
 && \hspace{3cm}+ \mu_{22} \left(- h \sin\alpha   + H \cos \alpha  + v_2\right)^2 \biggr]  ~ ,
\ee
and the mass-squared mixing coefficients for  $S$-$H$ and $S$-$h$ mixing can respectively be extracted via
\be
\mu_{SH} \equiv \frac{\partial^2 V}{\partial S \partial H} \biggr|_{h, H, S = 0} =  \sin\alpha ( v_1 \mu_{11}  + v_2 \mu_{12} )  + \cos\alpha(        v_1 \mu_{12} + v_2 \mu_{22}     ) \nonumber \\
\mu_{Sh} \equiv \frac{\partial^2 V}{\partial S \partial h} \biggr|_{h, H, S = 0} = \cos\alpha ( v_1 \mu_{11}  + v_2 \mu_{12} )  - \sin\alpha(        v_1 \mu_{12} + v_2 \mu_{22}     ) 
\ee
For parameter choices that satisfy $\mu_{Sh} \ll  \mu_{SH}  \ll v_h $ in the $\tan \beta \gg 1$ limits, $S$ will mix predominantly with the $H$ state
whose couplings are leptophilic and proportional to charged lepton masses. Diagonalizing this additional mixing between $S$-$H$ thus yields 
the $S \bar \ell \ell$ interactions in \Eq{eq:lag}. Note that, as in the case of one SM Higgs doublet, a additional singlet scalar can mix through the super-renormalizable portal wihtout acquiring a vacuum expectation value provided that a suitably chosen linear $S$ term is included in the potential (see \cite{Batell:2012mj} for a discussion). 

\section{$Z \to \ell^+ \ell^- \varphi$ decays}
\label{sec:zdecay}

The amplitude for radiative $\varphi$ production in $Z \to \ell^+(p_1) \ell^-(p_2) \varphi(p_3)$ decays depicted in Fig. \ref{fig:feyn} can be written 
\be
 {\cal M} =  \frac{ g_\tau m_Z}{v_h}   \bar u(p_1)
\!\biggl[ 
\frac{ (\slashed{p_1} + \slashed{p_3}) + m_\tau}{ m_{13}^2 -m_\ell^2} \gamma^\mu (g_V  +  g_A \gamma^5)
+ \gamma^\mu (g_V \! + \! g_A \gamma^5)\frac{ (\slashed{p_2} + \slashed{p_3}) + m_\tau}{ m_{12}^2 -m_\ell^2} 
\biggr] v(p_2),
\ee
where $p_i$ are the final state momenta with
$m_{ij}^2 \equiv (p_i + p_j)^2$,
$v_h = 246$ GeV is the SM Higgs vacuum expectation
value, and $g_{V,A}$ are respectively the tau vector and axial-vector charges, which satisfy
\be
g_V = -\frac{1}{2} (1 - 4 \sin^2\theta_W)~~,~~ g_A = -\frac{1}{2}~.
\ee
The partial width can be written 
\be
\label{eq:Zwidth}
\Gamma(Z\to \ell^+\ell^-\varphi) = 
\frac{1}{256\pi^3 m^3_Z} \int dm^2_{12}\int dm^2_{23} \, \overline{|{\cal M}|^2 }~,
\ee
where the $m^2_{12}$ integration limits satisfy
$
(m^2_{12})_{\rm min} = 4 m^2_\ell~,~  (m^2_{12})_{\rm max} = (m_Z-m_\varphi)^2~,$ 
and the $m^2_{23}$ limits are 
\be
(m^2_{23})_{\rm min}^{\rm max} = (E_2^* + E_3^*)^2 -\! \biggl( \!\sqrt{E^{*2}_2 - m_\ell^2}  \mp    
\sqrt{E^{*2}_3 - m_\varphi^2}\biggr)^{\! 2} ,
\ee
and we have defined
\be
E^*_{2} = \frac{m_{12}}{2} ~,~ E_3^* = \frac{ m_Z^2 - m_{12}^2 - m_\varphi^2 }{2m_{12}}~.
\ee


\section{Estimates on Acceptance Efficiency for $Z\rightarrow\tau\tau\varphi$}
\label{sec:taus}
%
\begin{figure*}[t]
    \includegraphics[width=3.2 in]{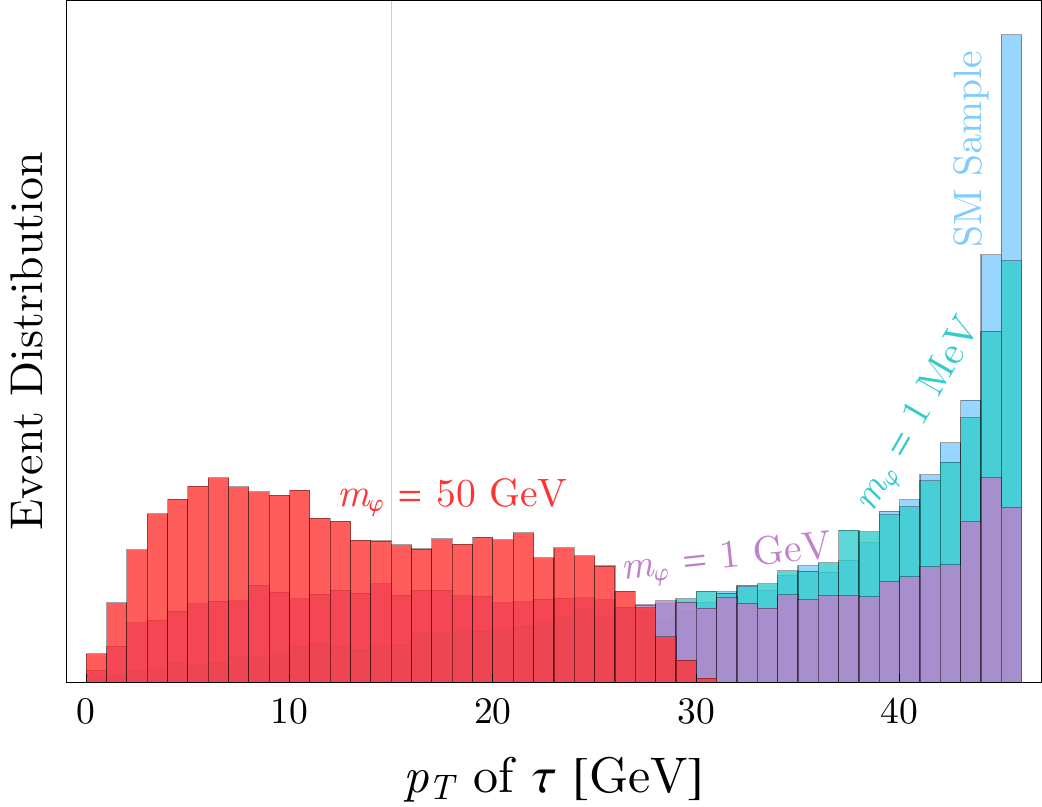} \qquad
    \includegraphics[width=3.2 in]{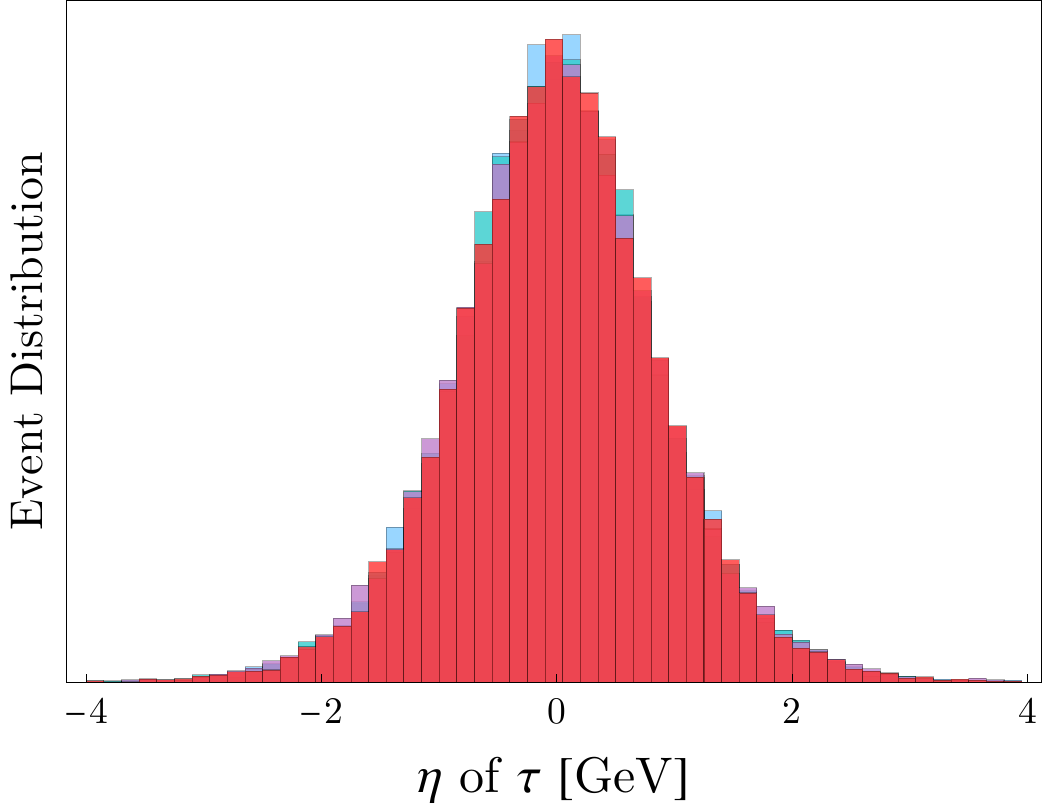} 
    \caption{The truth-level $p_T$ and $\eta$ distributions of the $\tau$ particles produced from $Z$ decays as a function of the $\varphi$ mass. }
    \label{fig:kinematics}
\end{figure*}
Since$\tau$decays are too complex to simply reconstruct the \textit{Z} peak, as a proxy for the much more involved analysis we perform truth-level cuts on taus from \texttt{MadGraph}-simulated events \cite{Alwall:2014hca} to estimate the number of new physics events that would resemble SM events close enough to be included in the decay width measurement.
As done in LEP and LHC \cite{ALEPH:2005ab, CMS:2018lkr} analyses, we put a cut on the transverse momentum of taus $p_T > 15$ GeV and on the total energy squared $s'$ of the $\tau\tau$ system to be $s' > 4 m_\tau^2$. 
While a more rigorous study would include evaluating the cuts of individual $\tau$ decay channels specifically and simulating the full $\tau$ decay, comparing truth-level $\tau$ information with and without the $\phi$ production gives us an order-of-magnitude understanding of how many new-physics events can be mistaken for SM events. 

We show the $\tau$ distributions in $p_T$ and $\eta$ in Figure~\ref{fig:kinematics}. 
Note that as the mass of the $\varphi$ is increased to $\mathcal{O}(m_Z \gtrsim 10 \text{ GeV})$, the $p_T$ of the taus is dramatically affected and most of the new physics events will be rejected. 
However, the $\eta$ distribution of the taus is largely unchanging.

\bibliographystyle{utphys3}
\bibliography{biblio}

\providecommand{\href}[2]{#2}\begingroup\raggedright\begin{thebibliography}{10}

\bibitem{Jungman:1995df}
G.~Jungman, M.~Kamionkowski, and K.~Griest, ``{Supersymmetric dark matter},''
  \href{https://dx.doi.org/10.1016/0370-1573(95)00058-5}{{\em Phys. Rept.}
  {\bfseries 267} (1996) 195--373},
  \href{https://arxiv.org/abs/hep-ph/9506380}{{\ttfamily
  arXiv:hep-ph/9506380}}.

\bibitem{Fox:2011fx}
P.~J. Fox, R.~Harnik, J.~Kopp, and Y.~Tsai, ``{LEP Shines Light on Dark
  Matter},'' \href{https://dx.doi.org/10.1103/PhysRevD.84.014028}{{\em Phys.
  Rev. D} {\bfseries 84} (2011) 014028},
  \href{https://arxiv.org/abs/1103.0240}{{\ttfamily arXiv:1103.0240 [hep-ph]}}.

\bibitem{Bai:2010hh}
Y.~Bai, P.~J. Fox, and R.~Harnik, ``{The Tevatron at the Frontier of Dark
  Matter Direct Detection},''
  \href{https://dx.doi.org/10.1007/JHEP12(2010)048}{{\em JHEP} {\bfseries 12}
  (2010) 048}, \href{https://arxiv.org/abs/1005.3797}{{\ttfamily
  arXiv:1005.3797 [hep-ph]}}.

\bibitem{Kahlhoefer:2017dnp}
F.~Kahlhoefer, ``{Review of LHC Dark Matter Searches},''
  \href{https://dx.doi.org/10.1142/S0217751X1730006X}{{\em Int. J. Mod. Phys.
  A} {\bfseries 32} no.~13, (2017) 1730006},
  \href{https://arxiv.org/abs/1702.02430}{{\ttfamily arXiv:1702.02430
  [hep-ph]}}.

\bibitem{Dreiner:2012xm}
H.~Dreiner, M.~Huck, M.~Kr\"amer, D.~Schmeier, and J.~Tattersall,
  ``{Illuminating Dark Matter at the ILC},''
  \href{https://dx.doi.org/10.1103/PhysRevD.87.075015}{{\em Phys. Rev. D}
  {\bfseries 87} no.~7, (2013) 075015},
  \href{https://arxiv.org/abs/1211.2254}{{\ttfamily arXiv:1211.2254 [hep-ph]}}.

\bibitem{Bernardi:2022hny}
G.~Bernardi {\em et~al.}, ``{The Future Circular Collider: a Summary for the US
  2021 Snowmass Process},'' \href{https://arxiv.org/abs/2203.06520}{{\ttfamily
  arXiv:2203.06520 [hep-ex]}}.

\bibitem{Han:2018wus}
T.~Han, S.~Mukhopadhyay, and X.~Wang, ``{Electroweak Dark Matter at Future
  Hadron Colliders},''
  \href{https://dx.doi.org/10.1103/PhysRevD.98.035026}{{\em Phys. Rev. D}
  {\bfseries 98} no.~3, (2018) 035026},
  \href{https://arxiv.org/abs/1805.00015}{{\ttfamily arXiv:1805.00015
  [hep-ph]}}.

\bibitem{Blaising:2021vhh}
{\bfseries CLICdp} Collaboration, J.-J. Blaising, P.~Roloff, A.~Sailer, and
  U.~Schnoor, ``{Physics performance for Dark Matter searches at $\sqrt{s}=$ 3
  TeV at CLIC using mono-photons and polarised beams},''
  \href{https://arxiv.org/abs/2103.06006}{{\ttfamily arXiv:2103.06006
  [hep-ex]}}.

\bibitem{Bai:2021rdg}
M.~Bai {\em et~al.}, ``{C$^3$: A ''Cool'' Route to the Higgs Boson and
  Beyond},'' in {\em {Snowmass 2021}}.
\newblock 10, 2021.
\newblock \href{https://arxiv.org/abs/2110.15800}{{\ttfamily arXiv:2110.15800
  [hep-ex]}}.

\bibitem{Liu:2019ogn}
Z.~Liu, Y.-H. Xu, and Y.~Zhang, ``{Probing dark matter particles at CEPC},''
  \href{https://dx.doi.org/10.1007/JHEP06(2019)009}{{\em JHEP} {\bfseries 06}
  (2019) 009}, \href{https://arxiv.org/abs/1903.12114}{{\ttfamily
  arXiv:1903.12114 [hep-ph]}}.

\bibitem{Cepeda:2019klc}
M.~Cepeda {\em et~al.}, ``{Report from Working Group 2}: {Higgs Physics at the
  HL-LHC and HE-LHC},''
  \href{https://dx.doi.org/10.23731/CYRM-2019-007.221}{{\em CERN Yellow Rep.
  Monogr.} {\bfseries 7} (2019) 221--584},
  \href{https://arxiv.org/abs/1902.00134}{{\ttfamily arXiv:1902.00134
  [hep-ph]}}.

\bibitem{AlAli:2021let}
H.~Al~Ali {\em et~al.}, ``{The muon Smasher\textquoteright{}s guide},''
  \href{https://dx.doi.org/10.1088/1361-6633/ac6678}{{\em Rept. Prog. Phys.}
  {\bfseries 85} no.~8, (2022) 084201},
  \href{https://arxiv.org/abs/2103.14043}{{\ttfamily arXiv:2103.14043
  [hep-ph]}}.

\bibitem{Han:2020uak}
T.~Han, Z.~Liu, L.-T. Wang, and X.~Wang, ``{WIMPs at High Energy Muon
  Colliders},'' \href{https://dx.doi.org/10.1103/PhysRevD.103.075004}{{\em
  Phys. Rev. D} {\bfseries 103} no.~7, (2021) 075004},
  \href{https://arxiv.org/abs/2009.11287}{{\ttfamily arXiv:2009.11287
  [hep-ph]}}.

\bibitem{Han:2022ubw}
T.~Han, Z.~Liu, L.-T. Wang, and X.~Wang, ``{WIMP Dark Matter at High Energy
  Muon Colliders $-$A White Paper for Snowmass 2021},'' in {\em {Snowmass
  2021}}.
\newblock 3, 2022.
\newblock \href{https://arxiv.org/abs/2203.07351}{{\ttfamily arXiv:2203.07351
  [hep-ph]}}.

\bibitem{Bottaro:2021snn}
S.~Bottaro, D.~Buttazzo, M.~Costa, R.~Franceschini, P.~Panci, D.~Redigolo, and
  L.~Vittorio, ``{Closing the window on WIMP Dark Matter},''
  \href{https://dx.doi.org/10.1140/epjc/s10052-021-09917-9}{{\em Eur. Phys. J.
  C} {\bfseries 82} no.~1, (2022) 31},
  \href{https://arxiv.org/abs/2107.09688}{{\ttfamily arXiv:2107.09688
  [hep-ph]}}.

\bibitem{Bottaro:2022one}
S.~Bottaro, D.~Buttazzo, M.~Costa, R.~Franceschini, P.~Panci, D.~Redigolo, and
  L.~Vittorio, ``{The last complex WIMPs standing},''
  \href{https://dx.doi.org/10.1140/epjc/s10052-022-10918-5}{{\em Eur. Phys. J.
  C} {\bfseries 82} no.~11, (2022) 992},
  \href{https://arxiv.org/abs/2205.04486}{{\ttfamily arXiv:2205.04486
  [hep-ph]}}.

\bibitem{Muong-2:2023cdq}
{\bfseries Muon g-2} Collaboration, D.~P. Aguillard {\em et~al.},
  ``{Measurement of the Positive Muon Anomalous Magnetic Moment to 0.20~ppm},''
  \href{https://dx.doi.org/10.1103/PhysRevLett.131.161802}{{\em Phys. Rev.
  Lett.} {\bfseries 131} no.~16, (2023) 161802},
  \href{https://arxiv.org/abs/2308.06230}{{\ttfamily arXiv:2308.06230
  [hep-ex]}}.

\bibitem{DAMIC-M:2023gxo}
{\bfseries DAMIC-M} Collaboration, I.~Arnquist {\em et~al.}, ``{First
  Constraints from DAMIC-M on Sub-GeV Dark-Matter Particles Interacting with
  Electrons},'' \href{https://dx.doi.org/10.1103/PhysRevLett.130.171003}{{\em
  Phys. Rev. Lett.} {\bfseries 130} no.~17, (2023) 171003},
  \href{https://arxiv.org/abs/2302.02372}{{\ttfamily arXiv:2302.02372
  [hep-ex]}}.

\bibitem{XENON:2019gfn}
{\bfseries XENON} Collaboration, E.~Aprile {\em et~al.}, ``{Light Dark Matter
  Search with Ionization Signals in XENON1T},''
  \href{https://dx.doi.org/10.1103/PhysRevLett.123.251801}{{\em Phys. Rev.
  Lett.} {\bfseries 123} no.~25, (2019) 251801},
  \href{https://arxiv.org/abs/1907.11485}{{\ttfamily arXiv:1907.11485
  [hep-ex]}}.

\bibitem{XENON:2021qze}
{\bfseries XENON} Collaboration, E.~Aprile {\em et~al.}, ``{Emission of single
  and few electrons in XENON1T and limits on light dark matter},''
  \href{https://dx.doi.org/10.1103/PhysRevD.106.022001}{{\em Phys. Rev. D}
  {\bfseries 106} no.~2, (2022) 022001},
  \href{https://arxiv.org/abs/2112.12116}{{\ttfamily arXiv:2112.12116
  [hep-ex]}}.

\bibitem{Drees:2001xw}
J.~Drees, ``{Review of final LEP results, or, A Tribute to LEP},''
  \href{https://dx.doi.org/10.1142/S0217751X02012727}{{\em Int. J. Mod. Phys.
  A} {\bfseries 17} (2002) 3259--3283},
  \href{https://arxiv.org/abs/hep-ex/0110077}{{\ttfamily
  arXiv:hep-ex/0110077}}.

\bibitem{Belle-II:2022yaw}
{\bfseries Belle-II} Collaboration, I.~Adachi {\em et~al.}, ``{Search for an
  Invisible Z' in a Final State with Two Muons and Missing Energy at Belle
  II},'' \href{https://dx.doi.org/10.1103/PhysRevLett.130.231801}{{\em Phys.
  Rev. Lett.} {\bfseries 130} no.~23, (2023) 231801},
  \href{https://arxiv.org/abs/2212.03066}{{\ttfamily arXiv:2212.03066
  [hep-ex]}}.

\bibitem{Lin:2013sca}
T.~Lin, E.~W. Kolb, and L.-T. Wang, ``{Probing dark matter couplings to top and
  bottom quarks at the LHC},''
  \href{https://dx.doi.org/10.1103/PhysRevD.88.063510}{{\em Phys. Rev. D}
  {\bfseries 88} no.~6, (2013) 063510},
  \href{https://arxiv.org/abs/1303.6638}{{\ttfamily arXiv:1303.6638 [hep-ph]}}.

\bibitem{Izaguirre:2014vva}
E.~Izaguirre, G.~Krnjaic, and B.~Shuve, ``{The Galactic Center Excess from the
  Bottom Up},'' \href{https://dx.doi.org/10.1103/PhysRevD.90.055002}{{\em Phys.
  Rev. D} {\bfseries 90} no.~5, (2014) 055002},
  \href{https://arxiv.org/abs/1404.2018}{{\ttfamily arXiv:1404.2018 [hep-ph]}}.

\bibitem{Cirigliano:2005ck}
V.~Cirigliano, B.~Grinstein, G.~Isidori, and M.~B. Wise, ``{Minimal flavor
  violation in the lepton sector},''
  \href{https://dx.doi.org/10.1016/j.nuclphysb.2005.08.037}{{\em Nucl. Phys. B}
  {\bfseries 728} (2005) 121--134},
  \href{https://arxiv.org/abs/hep-ph/0507001}{{\ttfamily
  arXiv:hep-ph/0507001}}.

\bibitem{Planck:2018vyg}
{\bfseries Planck} Collaboration, N.~Aghanim {\em et~al.}, ``{Planck 2018
  results. VI. Cosmological parameters},''
  \href{https://dx.doi.org/10.1051/0004-6361/201833910}{{\em Astron.
  Astrophys.} {\bfseries 641} (2020) A6},
  \href{https://arxiv.org/abs/1807.06209}{{\ttfamily arXiv:1807.06209
  [astro-ph.CO]}}. [Erratum: Astron.Astrophys. 652, C4 (2021)].

\bibitem{Craig:2013hca}
N.~Craig, J.~Galloway, and S.~Thomas, ``{Searching for Signs of the Second
  Higgs Doublet},'' \href{https://arxiv.org/abs/1305.2424}{{\ttfamily
  arXiv:1305.2424 [hep-ph]}}.

\bibitem{Krnjaic:2015mbs}
G.~Krnjaic, ``{Probing Light Thermal Dark-Matter With a Higgs Portal
  Mediator},'' \href{https://dx.doi.org/10.1103/PhysRevD.94.073009}{{\em Phys.
  Rev. D} {\bfseries 94} no.~7, (2016) 073009},
  \href{https://arxiv.org/abs/1512.04119}{{\ttfamily arXiv:1512.04119
  [hep-ph]}}.

\bibitem{Gondolo:1990dk}
P.~Gondolo and G.~Gelmini, ``{Cosmic abundances of stable particles: Improved
  analysis},'' \href{https://dx.doi.org/10.1016/0550-3213(91)90438-4}{{\em
  Nucl. Phys. B} {\bfseries 360} (1991) 145--179}.

\bibitem{Izaguirre:2015yja}
E.~Izaguirre, G.~Krnjaic, P.~Schuster, and N.~Toro, ``{Analyzing the Discovery
  Potential for Light Dark Matter},''
  \href{https://dx.doi.org/10.1103/PhysRevLett.115.251301}{{\em Phys. Rev.
  Lett.} {\bfseries 115} no.~25, (2015) 251301},
  \href{https://arxiv.org/abs/1505.00011}{{\ttfamily arXiv:1505.00011
  [hep-ph]}}.

\bibitem{Berlin:2018bsc}
A.~Berlin, N.~Blinov, G.~Krnjaic, P.~Schuster, and N.~Toro, ``{Dark Matter,
  Millicharges, Axion and Scalar Particles, Gauge Bosons, and Other New Physics
  with LDMX},'' \href{https://dx.doi.org/10.1103/PhysRevD.99.075001}{{\em Phys.
  Rev. D} {\bfseries 99} no.~7, (2019) 075001},
  \href{https://arxiv.org/abs/1807.01730}{{\ttfamily arXiv:1807.01730
  [hep-ph]}}.

\bibitem{Fabbrichesi:2020wbt}
M.~Fabbrichesi, E.~Gabrielli, and G.~Lanfranchi, ``{The Dark Photon},''
  \href{https://arxiv.org/abs/2005.01515}{{\ttfamily arXiv:2005.01515
  [hep-ph]}}.

\bibitem{Alwall:2014hca}
J.~Alwall, R.~Frederix, S.~Frixione, V.~Hirschi, F.~Maltoni, O.~Mattelaer,
  H.~S. Shao, T.~Stelzer, P.~Torrielli, and M.~Zaro, ``{The automated
  computation of tree-level and next-to-leading order differential cross
  sections, and their matching to parton shower simulations},''
  \href{https://dx.doi.org/10.1007/JHEP07(2014)079}{{\em JHEP} {\bfseries 07}
  (2014) 079}, \href{https://arxiv.org/abs/1405.0301}{{\ttfamily
  arXiv:1405.0301 [hep-ph]}}.

\bibitem{BaBar:2020jma}
{\bfseries BaBar} Collaboration, J.~P. Lees {\em et~al.}, ``{Search for a Dark
  Leptophilic Scalar in $e^+e^-$ Collisions},''
  \href{https://dx.doi.org/10.1103/PhysRevLett.125.181801}{{\em Phys. Rev.
  Lett.} {\bfseries 125} no.~18, (2020) 181801},
  \href{https://arxiv.org/abs/2005.01885}{{\ttfamily arXiv:2005.01885
  [hep-ex]}}.

\bibitem{ALEPH:2005ab}
{\bfseries ALEPH, DELPHI, L3, OPAL, SLD, LEP Electroweak Working Group, SLD
  Electroweak Group, SLD Heavy Flavour Group} Collaboration, S.~Schael {\em
  et~al.}, ``{Precision electroweak measurements on the $Z$ resonance},''
  \href{https://dx.doi.org/10.1016/j.physrep.2005.12.006}{{\em Phys. Rept.}
  {\bfseries 427} (2006) 257--454},
  \href{https://arxiv.org/abs/hep-ex/0509008}{{\ttfamily
  arXiv:hep-ex/0509008}}.

\bibitem{ParticleDataGroup:2020ssz}
{\bfseries Particle Data Group} Collaboration, P.~A. Zyla {\em et~al.},
  ``{Review of Particle Physics},''
  \href{https://dx.doi.org/10.1093/ptep/ptaa104}{{\em PTEP} {\bfseries 2020}
  no.~8, (2020) 083C01}.

\bibitem{Agapov:2022bhm}
I.~Agapov {\em et~al.}, ``{Future Circular Lepton Collider FCC-ee: Overview and
  Status},'' in {\em {Snowmass 2021}}.
\newblock 3, 2022.
\newblock \href{https://arxiv.org/abs/2203.08310}{{\ttfamily arXiv:2203.08310
  [physics.acc-ph]}}.

\bibitem{FCC:2018byv}
{\bfseries FCC} Collaboration, A.~Abada {\em et~al.}, ``{FCC Physics
  Opportunities}: {Future Circular Collider Conceptual Design Report Volume
  1},'' \href{https://dx.doi.org/10.1140/epjc/s10052-019-6904-3}{{\em Eur.
  Phys. J. C} {\bfseries 79} no.~6, (2019) 474}.

\bibitem{FCC:2018vvp}
{\bfseries FCC} Collaboration, A.~Abada {\em et~al.}, ``{FCC-hh: The Hadron
  Collider}: {Future Circular Collider Conceptual Design Report Volume 3},''
  \href{https://dx.doi.org/10.1140/epjst/e2019-900087-0}{{\em Eur. Phys. J. ST}
  {\bfseries 228} no.~4, (2019) 755--1107}.

\bibitem{FCC:2018evy}
{\bfseries FCC} Collaboration, A.~Abada {\em et~al.}, ``{FCC-ee: The Lepton
  Collider}: {Future Circular Collider Conceptual Design Report Volume 2},''
  \href{https://dx.doi.org/10.1140/epjst/e2019-900045-4}{{\em Eur. Phys. J. ST}
  {\bfseries 228} no.~2, (2019) 261--623}.

\bibitem{Kamenik:2023hvi}
J.~F. Kamenik, A.~Korajac, M.~Szewc, M.~Tammaro, and J.~Zupan,
  ``{Flavor-violating Higgs and Z boson decays at a future circular lepton
  collider},'' \href{https://dx.doi.org/10.1103/PhysRevD.109.L011301}{{\em
  Phys. Rev. D} {\bfseries 109} no.~1, (2024) L011301},
  \href{https://arxiv.org/abs/2306.17520}{{\ttfamily arXiv:2306.17520
  [hep-ph]}}.

\bibitem{Ho:2022ipo}
T.~S.~M. Ho, X.-H. Jiang, T.~H. Kwok, L.~Li, and T.~Liu, ``{Testing Lepton
  Flavor Universality at Future $Z$ Factories},''
  \href{https://arxiv.org/abs/2212.02433}{{\ttfamily arXiv:2212.02433
  [hep-ph]}}.

\bibitem{Budker:1969cd}
G.~I. Budker, ``{Accelerators and colliding beams},'' {\em Conf. Proc. C}
  {\bfseries 690827} (1969) 33--39.

\bibitem{earlyMuon:1996}
V.~V. Parkhomchuk and A.~N. Skrinsky, ``{Ionization cooling: Physics and
  applications},'' \href{https://dx.doi.org/10.1063/1.49355}{{\em AIP
  Conference Proceedings} {\bfseries 352} no.~1, (01, 1996) 7--9},
  \href{https://arxiv.org/abs/https://pubs.aip.org/aip/acp/article-pdf/352/1/7/11394715/7\_1\_online.pdf}{{\ttfamily
  https://pubs.aip.org/aip/acp/article-pdf/352/1/7/11394715/7\_1\_online.pdf}}.
  \url{https://doi.org/10.1063/1.49355}.

\bibitem{osti:1156195}
D.~Neuffer, ``Principles and applications of muon cooling,''
  \href{https://dx.doi.org/10.2172/1156195}{{\em Part.Accel.} }.
  \url{https://www.osti.gov/biblio/1156195}.

\bibitem{Mohayai:2018rxn}
{\bfseries MICE} Collaboration, T.~A. Mohayai,
  \href{https://dx.doi.org/10.18429/JACoW-IPAC2018-FRXGBE3}{``{First
  Demonstration of Ionization Cooling in MICE},''} in {\em {9th International
  Particle Accelerator Conference}}.
\newblock 6, 2018.
\newblock \href{https://arxiv.org/abs/1806.01807}{{\ttfamily arXiv:1806.01807
  [physics.acc-ph]}}.

\bibitem{Delahaye:2014vvd}
J.-P. Delahaye {\em et~al.},
  \href{https://dx.doi.org/10.18429/JACoW-IPAC2014-WEZA02}{``{A Staged Muon
  Accelerator Facility For Neutrino and Collider Physics},''} in {\em {5th
  International Particle Accelerator Conference}}, p.~WEZA02.
\newblock 6, 2014.
\newblock \href{https://arxiv.org/abs/1502.01647}{{\ttfamily
  arXiv:1502.01647}}.

\bibitem{Boscolo:2018ytm}
M.~Boscolo, J.-P. Delahaye, and M.~Palmer, ``{The future prospects of muon
  colliders and neutrino factories},''
  \href{https://dx.doi.org/10.1142/9789811209604_0010}{{\em Rev. Accel. Sci.
  Tech.} {\bfseries 10} no.~01, (2019) 189--214},
  \href{https://arxiv.org/abs/1808.01858}{{\ttfamily arXiv:1808.01858
  [physics.acc-ph]}}.

\bibitem{Barger:1997yk}
V.~D. Barger, M.~S. Berger, J.~F. Gunion, and T.~Han, ``{Precision W boson and
  top quark mass determinations at a muon collider},''
  \href{https://dx.doi.org/10.1103/PhysRevD.56.1714}{{\em Phys. Rev. D}
  {\bfseries 56} (1997) 1714--1722},
  \href{https://arxiv.org/abs/hep-ph/9702334}{{\ttfamily
  arXiv:hep-ph/9702334}}.

\bibitem{Han:2020uid}
T.~Han, Y.~Ma, and K.~Xie, ``{High energy leptonic collisions and electroweak
  parton distribution functions},''
  \href{https://dx.doi.org/10.1103/PhysRevD.103.L031301}{{\em Phys. Rev. D}
  {\bfseries 103} no.~3, (2021) L031301},
  \href{https://arxiv.org/abs/2007.14300}{{\ttfamily arXiv:2007.14300
  [hep-ph]}}.

\bibitem{Han:2021lnp}
T.~Han, W.~Kilian, N.~Kreher, Y.~Ma, J.~Reuter, T.~Striegl, and K.~Xie,
  ``{Precision test of the muon-Higgs coupling at a high-energy muon
  collider},'' \href{https://dx.doi.org/10.1007/JHEP12(2021)162}{{\em JHEP}
  {\bfseries 12} (2021) 162},
  \href{https://arxiv.org/abs/2108.05362}{{\ttfamily arXiv:2108.05362
  [hep-ph]}}.

\bibitem{Buttazzo:2018qqp}
D.~Buttazzo, D.~Redigolo, F.~Sala, and A.~Tesi, ``{Fusing Vectors into Scalars
  at High Energy Lepton Colliders},''
  \href{https://dx.doi.org/10.1007/JHEP11(2018)144}{{\em JHEP} {\bfseries 11}
  (2018) 144}, \href{https://arxiv.org/abs/1807.04743}{{\ttfamily
  arXiv:1807.04743 [hep-ph]}}.

\bibitem{Buttazzo:2020uzc}
D.~Buttazzo, R.~Franceschini, and A.~Wulzer, ``{Two Paths Towards Precision at
  a Very High Energy Lepton Collider},''
  \href{https://dx.doi.org/10.1007/JHEP05(2021)219}{{\em JHEP} {\bfseries 05}
  (2021) 219}, \href{https://arxiv.org/abs/2012.11555}{{\ttfamily
  arXiv:2012.11555 [hep-ph]}}.

\bibitem{Forslund:2023reu}
M.~Forslund and P.~Meade, ``{Precision Higgs width and couplings with a high
  energy muon collider},''
  \href{https://dx.doi.org/10.1007/JHEP01(2024)182}{{\em JHEP} {\bfseries 01}
  (2024) 182}, \href{https://arxiv.org/abs/2308.02633}{{\ttfamily
  arXiv:2308.02633 [hep-ph]}}.

\bibitem{Ruhdorfer:2023uea}
M.~Ruhdorfer, E.~Salvioni, and A.~Wulzer, ``{Invisible Higgs boson decay from
  forward muons at a muon collider},''
  \href{https://dx.doi.org/10.1103/PhysRevD.107.095038}{{\em Phys. Rev. D}
  {\bfseries 107} no.~9, (2023) 095038},
  \href{https://arxiv.org/abs/2303.14202}{{\ttfamily arXiv:2303.14202
  [hep-ph]}}.

\bibitem{Accettura:2023ked}
C.~Accettura {\em et~al.}, ``{Towards a muon collider},''
  \href{https://dx.doi.org/10.1140/epjc/s10052-023-11889-x}{{\em Eur. Phys. J.
  C} {\bfseries 83} no.~9, (2023) 864},
  \href{https://arxiv.org/abs/2303.08533}{{\ttfamily arXiv:2303.08533
  [physics.acc-ph]}}. [Erratum: Eur.Phys.J.C 84, 36 (2024)].

\bibitem{Capdevilla:2021kcf}
R.~Capdevilla, D.~Curtin, Y.~Kahn, and G.~Krnjaic, ``{Systematically testing
  singlet models for (g - 2)},''
  \href{https://dx.doi.org/10.1007/JHEP04(2022)129}{{\em JHEP} {\bfseries 04}
  (2022) 129}, \href{https://arxiv.org/abs/2112.08377}{{\ttfamily
  arXiv:2112.08377 [hep-ph]}}.

\bibitem{Muong-2:2006rrc}
{\bfseries Muon g-2} Collaboration, G.~W. Bennett {\em et~al.}, ``{Final Report
  of the Muon E821 Anomalous Magnetic Moment Measurement at BNL},''
  \href{https://dx.doi.org/10.1103/PhysRevD.73.072003}{{\em Phys. Rev. D}
  {\bfseries 73} (2006) 072003},
  \href{https://arxiv.org/abs/hep-ex/0602035}{{\ttfamily
  arXiv:hep-ex/0602035}}.

\bibitem{Aoyama:2020ynm}
T.~Aoyama {\em et~al.}, ``{The anomalous magnetic moment of the muon in the
  Standard Model},''
  \href{https://dx.doi.org/10.1016/j.physrep.2020.07.006}{{\em Phys. Rept.}
  {\bfseries 887} (2020) 1--166},
  \href{https://arxiv.org/abs/2006.04822}{{\ttfamily arXiv:2006.04822
  [hep-ph]}}.

\bibitem{CMD-3:2023alj}
{\bfseries CMD-3} Collaboration, F.~V. Ignatov {\em et~al.}, ``{Measurement of
  the $e^+e^-\to\pi^+\pi^-$ cross section from threshold to 1.2 GeV with the
  CMD-3 detector},'' \href{https://arxiv.org/abs/2302.08834}{{\ttfamily
  arXiv:2302.08834 [hep-ex]}}.

\bibitem{Krnjaic:2022ozp}
G.~Krnjaic {\em et~al.}, ``{A Snowmass Whitepaper: Dark Matter Production at
  Intensity-Frontier Experiments},''
  \href{https://arxiv.org/abs/2207.00597}{{\ttfamily arXiv:2207.00597
  [hep-ph]}}.

\bibitem{Kahn:2018cqs}
Y.~Kahn, G.~Krnjaic, N.~Tran, and A.~Whitbeck, ``{M$^{3}$: a new muon missing
  momentum experiment to probe (g -2) and dark matter at Fermilab},''
  \href{https://dx.doi.org/10.1007/JHEP09(2018)153}{{\em JHEP} {\bfseries 09}
  (2018) 153}, \href{https://arxiv.org/abs/1804.03144}{{\ttfamily
  arXiv:1804.03144 [hep-ph]}}.

\bibitem{Andreev:2024sgn}
Y.~M. Andreev {\em et~al.}, ``{Exploration of the Muon g-2 and Light Dark
  Matter explanations in NA64 with the CERN SPS high energy muon beam},''
  \href{https://arxiv.org/abs/2401.01708}{{\ttfamily arXiv:2401.01708
  [hep-ex]}}.

\bibitem{Shifman:1978zn}
M.~A. Shifman, A.~I. Vainshtein, and V.~I. Zakharov, ``{Remarks on Higgs Boson
  Interactions with Nucleons},''
  \href{https://dx.doi.org/10.1016/0370-2693(78)90481-1}{{\em Phys. Lett. B}
  {\bfseries 78} (1978) 443--446}.

\bibitem{Cirelli:2013ufw}
M.~Cirelli, E.~Del~Nobile, and P.~Panci, ``{Tools for model-independent bounds
  in direct dark matter searches},''
  \href{https://dx.doi.org/10.1088/1475-7516/2013/10/019}{{\em JCAP} {\bfseries
  10} (2013) 019}, \href{https://arxiv.org/abs/1307.5955}{{\ttfamily
  arXiv:1307.5955 [hep-ph]}}.

\bibitem{XENON:2023cxc}
{\bfseries XENON} Collaboration, E.~Aprile {\em et~al.}, ``{First Dark Matter
  Search with Nuclear Recoils from the XENONnT Experiment},''
  \href{https://dx.doi.org/10.1103/PhysRevLett.131.041003}{{\em Phys. Rev.
  Lett.} {\bfseries 131} no.~4, (2023) 041003},
  \href{https://arxiv.org/abs/2303.14729}{{\ttfamily arXiv:2303.14729
  [hep-ex]}}.

\bibitem{Boehm:2013jpa}
C.~Boehm, M.~J. Dolan, and C.~McCabe, ``{A Lower Bound on the Mass of Cold
  Thermal Dark Matter from Planck},''
  \href{https://dx.doi.org/10.1088/1475-7516/2013/08/041}{{\em JCAP} {\bfseries
  08} (2013) 041}, \href{https://arxiv.org/abs/1303.6270}{{\ttfamily
  arXiv:1303.6270 [hep-ph]}}.

\bibitem{Nollett:2013pwa}
K.~M. Nollett and G.~Steigman, ``{BBN And The CMB Constrain Light,
  Electromagnetically Coupled WIMPs},''
  \href{https://dx.doi.org/10.1103/PhysRevD.89.083508}{{\em Phys. Rev. D}
  {\bfseries 89} no.~8, (2014) 083508},
  \href{https://arxiv.org/abs/1312.5725}{{\ttfamily arXiv:1312.5725
  [astro-ph.CO]}}.

\bibitem{Krnjaic:2019dzc}
G.~Krnjaic and S.~D. McDermott, ``{Implications of BBN Bounds for Cosmic Ray
  Upscattered Dark Matter},''
  \href{https://dx.doi.org/10.1103/PhysRevD.101.123022}{{\em Phys. Rev. D}
  {\bfseries 101} no.~12, (2020) 123022},
  \href{https://arxiv.org/abs/1908.00007}{{\ttfamily arXiv:1908.00007
  [hep-ph]}}.

\bibitem{Finkbeiner_2012}
D.~P. Finkbeiner, S.~Galli, T.~Lin, and T.~R. Slatyer, ``Searching for dark
  matter in the {CMB}: A compact parametrization of energy injection from new
  physics,'' \href{https://dx.doi.org/10.1103/physrevd.85.043522}{{\em Physical
  Review D} {\bfseries 85} no.~4, (Feb, 2012) }.
  \url{https://doi.org/10.1103%2Fphysrevd.85.043522}.

\bibitem{Batell:2012mj}
B.~Batell, D.~McKeen, and M.~Pospelov, ``{Singlet Neighbors of the Higgs
  Boson},'' \href{https://dx.doi.org/10.1007/JHEP10(2012)104}{{\em JHEP}
  {\bfseries 10} (2012) 104}, \href{https://arxiv.org/abs/1207.6252}{{\ttfamily
  arXiv:1207.6252 [hep-ph]}}.

\bibitem{CMS:2018lkr}
{\bfseries CMS} Collaboration, A.~M. Sirunyan {\em et~al.}, ``{Measurement of
  the $\mathrm{Z}\gamma^{*} \to \tau\tau$ cross section in pp collisions at
  $\sqrt{s} = $ 13 TeV and validation of $\tau$ lepton analysis techniques},''
  \href{https://dx.doi.org/10.1140/epjc/s10052-018-6146-9}{{\em Eur. Phys. J.
  C} {\bfseries 78} no.~9, (2018) 708},
  \href{https://arxiv.org/abs/1801.03535}{{\ttfamily arXiv:1801.03535
  [hep-ex]}}.

\end{thebibliography}\endgroup

\end{document}